**Evidence for Reduced Sensory Precision and Increased Reliance on Priors in Hallucination-Prone Individuals in a General Population Sample**

David Benrimoh[1†], Victoria L. Fisher[2†], Rashina Seabury[2], Ely Sibarium[2], Catalina Mourgues[2], Doris Chen[2], Albert Powers[2]

[†]Contributed equally to this manuscript

1. McGill University School of Medicine, Montreal, Canada

2. Yale University School of Medicine and the Connecticut Mental Health Center, New Haven, CT, USA

*Correspondence should be addressed to:

Albert R. Powers, M.D., Ph.D.
The Connecticut Mental Health Center, Rm. S109
34 Park Street
New Haven, CT 06519
albert.powers@yale.edu
203.974.7329

**Abstract**

**Background:** There is increasing evidence that people with hallucinations overweight perceptual beliefs relative to incoming sensory evidence. Much past work demonstrating prior overweighting has used simple, non-linguistic stimuli. However, auditory hallucinations in psychosis are often complex and linguistic. There may be an interaction between the type of auditory information being processed and its perceived quality in engendering hallucinations.

**Study Design:** We administered a linguistic version of the Conditioned Hallucinations (CH) task to an online sample of 88 general population participants. Metrics related to hallucination-proneness, recent auditory hallucinations, stimulus thresholds, and stimulus detection were collected; data was used to fit parameters of a Hierarchical Gaussian Filter (HGF) model of perceptual inference to determine how latent perceptual states influenced task behavior.

**Study Results:** Replicating past results, higher CH rates were associated with measures of higher hallucination-proneness and recent hallucinatory experiences; CH rates were positively correlated with increased prior weighting; and increased prior weighting was related to recent hallucinatory experiences. Unlike past results, participants with recent hallucinatory experiences as well as those with higher hallucination-proneness had higher stimulus thresholds, lower sensitivity to stimuli presented at the highest threshold, and tended to have lower response confidence, consistent with lower precision of sensory evidence.

**Conclusions:** We demonstrate that hallucination-prone individuals in the general population have increased conditioned hallucination rates using a linguistic version of the CH task, and replicated the finding that increased CH rates and recent hallucinations correlate with increased prior weighting. Results support a role for reduced sensory precision in the interplay between prior weighting and hallucination-proneness.

## Introduction

Computational techniques have increasingly been used in recent years to characterize cognitive and perceptual mechanisms that may underlie psychotic symptoms, with the goal of facilitating the development of novel treatments[1–8]. One finding that has gained prominence is the importance of the overweighting of prior beliefs in the generation of positive symptoms, such as hallucinations [1,2,5,8–15]. Previous work demonstrates that perceptual systems, rather than relying directly upon input from the sensory organs, use sensory input to update probabilistic models of the causes of sensations (the environment)[11,16,17]. In this view, perception is an inferential process in which organisms infer what is around them by combining sensory input with their prior beliefs about the world, weighted by the reliability of these different information channels. This weighted blending of sensory input and priors can be observed in many common situations (e.g., the use of lip-reading cues[18] and sentence context[19] in understanding speech in noisy environments; the use of shading to infer three-dimensional surfaces in the visual domain[20]). Bayesian statistical models have been used to investigate this combination of sensory input and prior knowledge[17]; these models have succeeded in predicting performance on a number of perceptual tasks[21–23] as well as neural activity in sensory contexts[24]. Substantial recent work has established that, at least in some cases, overreliance on priors compared to incoming sensory evidence may underpin hallucinations[5]. While evidence has pointed to this overweighting of priors, what is less clear is precisely how this takes place: is the relative overweighting of priors seen in hallucinations independent of or related to the absolute weighting (and quality) of incoming sensory information [1,2,8,25]?

This question is by no means purely academic; understanding how prior overweighting and resultant hallucinations relate to upstream sensory processing errors could inform different interventions based on identification of different distal mechanisms. For example, *in silico* simulations[1,2] demonstrated that an overweighting of certain priors can generate hallucinations, though this only occurs in the setting of permissively low sensory precision. Benrimoh et al. (2018, 2019) theorized that maladaptive priors would be encoded in upper levels of the processing hierarchy with weighting of some priors modified by dopaminergic signaling; whereas lower-quality sensory evidence could be due to reduced integrity of white matter connections such as the arcuate fasciculus (which is often degraded in schizophrenia[26]) or encoded by changes in cholinergic tone, also known to be altered in schizophrenia[27]. This view of maladaptive priors being expressed as perceptual changes only when sensory precision is permissively low suggests a different mechanism from, for example, a scenario where there is no absolute change in prior weighting but only low weighting of sensory evidence arising from low-quality sensory information, resulting in a relative over-weighting of priors, used to fill in missing sensory data.

Here, we attempt to interrogate this question of relative prior and sensory weighting using the conditioned hallucinations (CH) task[5,12,13]. CH occur as a result of classical conditioning, where a subject is presented with a salient stimulus paired with a difficult-to-detect target (e.g., an image and a sound) at the same time in a repeated manner, such that in the presence of the salient stimulus (e.g., the image) and the absence of the target, the subject may hallucinate the target in order to satisfy their expectation of the stimulus being present. Our group has developed and validated several versions of this task, which has been shown to be sensitive to susceptibility to psychosis and hallucinations[5,12] and current hallucination state[13]. However, this task has not yet been administered in a large, general-population sample, which would allow for a nuanced assessment of the interplay between sensory processing and hallucination-proneness outside of frank psychopathology or the presence of frank hallucinations.

**Methods**

*Participants*

Participants were recruited and completed the experimental task using Amazon Mechanical Turk. Based on a power analysis using initial pilot data, a sample size of 90 was determined to be appropriate to detect differences in computational parameters derived from participant responses. Pilot data were not included in the final analysis. In order to be included in the analysis, participants needed to report the absence of a current or past psychotic illness. It is important to note that this was *not* a healthy control sample, but rather a non-psychotic general population sample. This was necessary because removing participants with any mental illness may have artificially reduced the number of participants who demonstrate hallucination-proneness, given that there is some evidence of more frequent psychopathology (such as anxious and depressive symptoms) in those with hallucination or delusion-proneness[28–30].

120 participants signed up to complete the task. Of these, 21 did not complete the task. Eight completed the task but provided data that did not pass quality control screening [13]. Two subjects endorsed having a medical condition that could cause hallucinations and their data were discarded. Similarly, one subject endorsed drug or alcohol use in the 2-3 hours preceding their participation and their data were discarded. This resulted in 88 subjects available for analysis. Within these 88 subjects, one reported being red-green color blind and three reported corrected-to-normal (rather than normal) hearing; post-hoc analysis revealed no significant differences between their median CH rates and the rest of the sample ($H_1$= 0.75, p = .39), and, as such, these subjects were included in final analysis.

*Measures*

Participants completed several questionnaires prior to the CH task. These included the PHQ-9 and GAD-7 to screen for depression and generalized anxiety symptoms, respectively[31,32]; the Launay-Slade Hallucination Scale (LSHS)[33] to measure hallucination-proneness; the Auditory Hallucination Rating Scale (AHRS), to determine the intensity and nature of any recent hallucinatory experiences[34]; the Adverse Childhood Events and Trauma History Questionnaire (THQ)[35] given the strong association between psychosis and trauma[36]; as well as demographic information and psychiatric history.

For analysis, the sample was split into groups based on hallucination-propensity, which was generally low (mean LSHS score: 7.1 out of 60; SD: 9.4; see **Table 1**). Participants were designated as having high hallucination-propensity (HP) if they scored at least one standard deviation above the mean on the LSHS (n = 15); all other participants were designated as low propensity (LP; n = 73). The AHRS, which focuses more on recent hallucinatory experiences, had a mean of 4.8 (SD = 7.2) out of 49 possible points, again demonstrating a low rate of hallucinatory experiences in this population. We also split the sample into subgroups based on recent hallucination status using their responses to the AHRS. Those with an AHRS score of 0 (i.e., with no significant recent hallucinatory experiences) were put into the 'no recent hallucinations' subgroup (AH-), and those with a score above 0 were put into the 'some recent hallucinatory experiences' subgroup (AH+); each subgroup had 44 members.

*Materials*

Participants used their own computers to complete the CH task (see below), and did so in their environment of choice. All participants were asked to use headphones, ensure that their keyboard was working, and keep their screen brightness and volume at a maximal level. Prior work has demonstrated that the combination of these hardware checks and thresholding (which uses participant responses rather than hard-wired stimulus intensities) results in insensitivity of task performance to hardware, operating system, and browser differences[13]. All stimuli included in the task were presented via React (https://reactjs.org/).

*Conditioned Hallucination (CH) Task*

Each participant completed the three different versions of the CH task in random order. This was done in order to gather pilot data for an affective manipulation (showing images prior to the task), which is not the focus of this analysis. Here, we focus on only the 'neutral' condition, where subjects saw only gray squares before the task started.

The linguistic CH task was adapted from past instantiations of the original CH task[5,12,13] which uses Pavlovian associated-learning to elicit hallucinations (**Fig 1a**). In the CH task, participants are instructed to report if they detected an auditory target embedded in 70 db-SPL white noise. The target is always paired with a salient visual cue. In the original task, the auditory target was a 1-kHz pure tone. In this version, participants were instructed to detect a voice that said a nonsense word. A verbal target stimulus was chosen to more closely mimic hallucinations in psychosis, as has been done previously in general-population samples[15].

Participants first completed a short practice to familiarize themselves with the CH task. Participants repeated the practice until they reached an accuracy of at least 70%, as in past work[5,13]. After this, individual threshold estimates were obtained using the maximum-likelihood-based QUEST procedure [5,37]. We then fit the 75% threshold to a standard psychometric curve to determine target volumes at which participants were likely to report detection at rates of 50% and 25%.

Consistent with prior work[5,12,13], participants completed 12 blocks of the target detection task at each of the thresholds in addition to no-voice trials during which only the visual pattern was presented. As past work has demonstrated the emergence of groupwise difference within the first half of the experiment[13], we halved the number of trials per block from 30 to 15 to minimize fatigue. In order to smoothly build the association between the auditory target and the accompanying visual stimulus, the probability of at-threshold trials decreased over the blocks and sub-threshold and target-absent trials increased (**Fig 1b**).

Each version of the task had a different pattern-word pairing: red vertical stripes and 'dob', green horizontal stripes and 'bez', or black and white checkerboard and 'yag'. This was done to avoid stimulus pairs learned in one version of the task influencing CH rates in a version of the task completed later. As in past work, CH were defined as the participant indicating the presence of a voice stimulus on a trial when no target was present (i.e., the no-target trials)[5,12,13]. The CH rate was then defined as the proportion of no-target trials in which the participant reported hearing a voice. We found no significant relationship between CH rates and stimuli used during the neutral condition or order of task versions (Stimuli: $H_2$ = 1.7, p = .43; Order: $H_2$ = .72, p = .69), QUEST-derived threshold (Stimuli: $H_2$ = 0.40, p = .82; Order: $H_2$ = 0.85, p = .65), AHRS subgroups (Stimuli: $\chi^2$ = 0.85, p = .65; Order: $\chi^2$ = 1.0, p = .60) or LSHS subgroups (Stimuli: $\chi^2$ = 3.1, p = .21; Order: $\chi^2$ = 0.36, p = .84 ).

In addition to reporting detection / non-detection of the target, participants were asked to hold down the response button to indicate their degree of confidence in this judgment, guided by a visual scale appearing at fixation and moving from 1 ("Unsure") to 5 ("Confident") as the response button was held down.

*Hierarchical Gaussian Filter (HGF) Analysis*

To pinpoint potential mechanisms driving behavioral differences associated with hallucination-proneness, we employed the HGF model, a hierarchical Bayesian model of learning in a dynamic environment (see **Fig. 4a**) [38,39]. A three-tiered version of the HGF has been adapted to use trial-wise response and stimulus intensity values from the CH task[5,12,13] Extensive details of this model have been previously published[5,12,13]. The model estimates belief states during the task. Level 1 ($\mu_1$) corresponds to the belief that there was a voice during a given presentation of the visual stimulus. Level 2 ($\mu_2$) corresponds to the belief that, in general, a voice is predicted by the presence of the visual stimulus. Level 3 ($\mu_3$) corresponds to belief in the volatility of the association between voice and the visual stimulus (i.e., beliefs about the volatility of the relationship captured by level 2). Critically, the model also allows the estimation of parameters reflecting the weight afforded to (i.e., the precision of) prior expectations ($\nu$), evolution rates in the strength of association ($\omega_2$), and volatility ($\omega_3$) beliefs. Finally, the model translates posterior beliefs in the presence of the tone on any given trial into a probability of reporting detection using a softmax function, the slope of which corresponds to inverse decision noise ($\beta^{-1}$). To implement the HGF, we used the TAPAS computational toolbox (github.com/translationalneuromodeling/tapas) which is freely available with all relevant code.

Previous work using the HGF with an online version of the CH task used the empirical grand mean detection rates instead of expected detection rates. We explored which version of the model best explained the data. We performed Bayesian model comparison using spm_BMS[40]. Model exceedance probabilities showed support for the expected means (PXP = 0.514) over the empirical (PXP = 0.486). Additionally, a paired t-test of the inversion results from each model indicated a significant increase in percent-identical responses for the expected (90.6%) over the empirical (90.4%; $t_{87}$ = 2.74, p = 7.30 x $10^{-3}$). Thus, we performed all analyses using the expected means as the ground-truths.

*Statistical Analyses*

We aimed to determine if performance on the CH task was sensitive to hallucinations proneness in this general population sample. Thus, we used regression to test the relationship between CH rates and total score of the LSHS and AHRS. We also tested for mean differences between Low- and High-proneness groups, and those reporting vs. not reporting recent hallucinatory experiences, as outlined below. Prior to analysis, participant response data were used to fit the parameters of a three-tier Hierarchical Gaussian Filter (HGF), described in detail in prior work [5,12,13]. This was done in order to estimate latent states driving behavior and behavioral differences between groups in the task. In past work, we have demonstrated that the parameter $\nu$, which represents the weighting of prior beliefs relative to that of sensory evidence during perception, as particularly important for conditioned and clinical hallucination propensity [5,12,13]. Exploratory analyses sought to compare groups on secondary measures of task performance and HGF parameter estimates.

Behavioral data were not normally distributed, being skewed towards a low CH rate; this was expected in a general population sample. As such, non-parametric statistical tests were used where possible. Kruskall-Wallace tests were used to examine differences between participant subgroups for each condition, and linear models were used to determine the relationship between hallucination-proneness and parameters of interest. Repeated measures ANOVA[41] was used to assess relationships between subgroups over trials and blocks, as there is no equivalent nonparametric test for three-way comparisons. We did not correct for multiple comparisons given the exploratory nature of this analysis. Analyses were carried out using R Version 4.1.0 [42] and Python Version 3.9.13.

In addition, exploratory analyses of metacognition were carried out (see supplementary material).

**Results**

*Sample Characteristics*

**Table 1** displays demographic and clinical characteristics of participants retained for analysis separated by hallucination-proneness and recent hallucinatory experiences subgroups. Groups did not differ by sex or race, with the population primarily composed of individuals identifying as male (n = 62; 70.5%) and Caucasian (n = 67; 76.1%). The mean age of the participants retained for analysis was 39.5 (SD: 11.2), with a range of 20-66. The high hallucination-proneness groups were significantly younger than the low-proneness groups (AH- and AH+: $T_{86}$ = -2.64, p = 0.01; Low- and High-Propensity: $T_{56}$ = -6.18, p = 7.60 x $10^{-8}$ ). 90% of the sample were participants based in the U.S., and all participants reported English as their primary spoken language. One participant endorsed regular cannabis use in the last four weeks, and two endorsed higher-than-recommended levels of alcohol consumption. No participants reported currently being under the influence of a substance. A total of five subjects endorsed a history of mental illness, including depression, social anxiety, and OCD (**Table S1**); one participant endorsed current antidepressant use but no other participants endorsed active use of medications.

While the majority of the sample did not endorse depression or anxiety, 18.2% did endorse clinically significant depressive symptoms (≥ moderate symptom severity on PHQ-9), and 11.3% endorsed significant anxiety (≥ moderate symptom severity on GAD-7). With respect to trauma, the mean number of traumatic events experienced was 1.6 (SD: 1.9) and the mean number of adverse events experienced specifically in childhood (ACE score) was 1.1 (SD: 1.7), with nine participants (10.2%) reporting an ACE score of 4 or more.

Hallucination proneness subgroups defined by both hallucination measures (LSHS and AHRS) did not differ in the proportion reporting psychiatric diagnosis or prescribed medications, but did differ in the general severity and presence of depression and anxiety symptoms. High-proneness groups endorsed higher clinically-relevant depressive (AH- and AH+: $t_{60}$ = 3.62, p = 6.03 x $10^{-3}$; Low- and High-Propensity: $t_{17}$ = 4.63, p = 2.44 x $10^{-3}$) and anxiety (AH- and AH+: $t_{62}$ = 2.8, p = 6.47 x $10^{-3}$; LSHS: $t_{17}$ = 3.22, p = 4.98 x $10^{-3}$) symptoms. Conversely, low-proneness groups endorsed greater recent trauma (AH- and AH+: $t_{80}$ ; Low- and High-Propensity: $t_{57}$ = -5.68, p = 4.69 x $10^{-7}$). Groups did not differ in either childhood trauma or delusional ideation. Together, results are consistent with previous studies of psychosis-proneness [29,43].

*Hallucination proneness is associated with poorer sensory performance and greater CH rates*

High-proneness groups had higher QUEST-derived thresholds than low-proneness groups (AH- and AH+: $H_1$ = 5.76, p = 0.016; Low- and High-Propensity: $H_1$ = 8.61, p = 3.33 x $10^{-3}$; **Fig 2a & 2d**), meaning that they required higher stimulus intensity levels in order to report detection. This relationship tracked with severity of both recent hallucinatory experience and hallucination proneness (AHRS total score: $F_{1,86}$ = 13.9, p = 3.42 x $10^{-4}$, R = .36; LSHS total score: $F_{1,86}$ = 12.7, p = 6.00 x $10^{-4}$, R = .39; **Fig. S1a and S1c**).

As observed in previous work[5,13], those in hallucination-prone groups were more likely to experience conditioned hallucinations (AH- and AH+: $H_1$ = 5.67, p = 0.017; Low- & High-Propensity: $H_1$ = 6.16, p = 0.013; **Fig 2b & 2e**). Also consistent with recent findings, these differences tracked more with severity of recent hallucinations ($F_{1,86}$ = 8.37, p = 4.83 x $10^{-3}$, R = .28; **Fig S1b**) than with general propensity ($F_{1,86}$ = 3.86, p = .05, R = .18; **Fig S1d**).

We next investigated the relationship between QUEST-derived detection thresholds and task performance. We did not find that threshold value *itself* was predictive of detection during 75% condition trials ($F_{1,86}$ = 2.19, p = .14) or CH rate ($F_{1,86}$ = .24, p = 0.63). However, we did find that, despite having higher threshold volumes, those with high general proneness were less likely to report detecting the tone on 75% conditions ($H_1$ = 4.93, p = .026; **Fig 2e**). We observed a significant negative relationship between detection at the 75% condition with both general ($F_{1,86}$ = 4.00, p = .049; **Fig S1d**) and recent ($F_{1,86}$ = 6.02, p = .016; **Fig S1b**) hallucination-proneness.

*Hallucination-prone groups are less confident in reporting detection on target-present trials and have reduced confidence when reporting non-detection in target-absent trials*

While AH+ and AH- groups did not differ in overall confidence levels **(Fig 2c & 2g)**, in our modeling of confidence we found significant interactions between responses (yes or no) and group as well as a three-way interaction between response, group, and condition (no voice, 25, 50, 75%; **Table S2**). Post-hoc analyses revealed that, as severity of recent hallucinations increased, confidence in detecting the target on the 75% condition ($F_{1,86}$ = 4.80, p = .03, R = -.20) and when reporting non-detection on the no-target ($F_{1,86}$ = 9.19, p = 2.26 x $10^{-3}$, R = -.30), 25% ($F_{1,86}$ = 6.34, p = .01, R = -.24) and 50% ($F_{1,85}$ = 4.02, p = .048, R = -.18) conditions decreased. We also explored this relationship using our general proneness metrics (Low- and High-Propensity groups, LSHS-total) using repeated measures ANOVA (subgroups; **Table S3**). We observed similar effects as AHRS grouping; however, there was a main effect of LSHS group on confidence. Post-hoc analyses indicated significantly decreased confidence in the High severity group when reporting yes on the 75% condition ($H_1$ = 5.70, p = .02) and reporting non-detection on the no-target ($H_1$ = 6.32, p = 0.012) , 25% ($H_1$ = 4.64, p = .03), and 50%($H_1$ = 4.03, p = .047) conditions. These effects also scaled with severity of symptoms, as regression analyses indicated significant negative relationships between LSHS total score and confidence with each condition (No-Voice, no: $F_{1,86}$ = 17.4, p = 7.31 x $10^{-5}$, R = -.40; 25%, no: $F_{1,86}$ = 11.93, p = 8.59 x $10^{-4}$, R = -.33; 50%, no: $F_{1,85}$ = 10.83, p = 1.45 x $10^{-3}$, R = -.32); 75%, yes: $F_{1,86}$ = 12.6, p = 6.29 x $10^{-4}$, R = -.34).

*High-proneness and high-severity groups maintain high CH rates over time*

We additionally explored how propensity for conditioned hallucinations changed throughout the course of the task (**Fig 3a** & **3e**). A repeated measures ANOVA (**Table S4 & S5**) revealed a significant effect of group (AH- and AH+: $F_{1,86}$ = 11.68, p = 9.67 x $10^{-3}$; Low- and High-Propensity: $F_{1,86}$ = 6.92, p = 0.010) and a group-by-trial interaction (AH- and AH+: $F_{1,6158}$ = 8.20, p = 4.2 x $10^{-3}$; $F_{1,6158}$ = 61.3, p = 5.74 x $10^{-15}$). A post-hoc analysis with both AH- and AH+ , Low- and High-Propensity and trial number in a single RMANOVA (**Table S6)** indicated that only AH- and AH+ groups had a significant effect on cumulative CH rate ($F_{1,86}$ = 11.68, p = 9.67 x $10^{-3}$). Further, while both AH- and AH+ and Low- and High-Propensity groups had significant interactions with trial type, the effect size for Low- and High-Propensity groups (Low- and High-Propensity: $\eta^2$ = 0.009, AH- and AH+: $\eta^2$ = 0.001) was larger.

We also explored how detection rates during the 75% condition changed throughout the experiment (**Fig 3d and 3h**). Repeated measures ANOVAs (**Table S7** & **Table S8**) indicated a significant effect of the interaction between hallucination proneness group and trial for both recent hallucinations ($F_{1,3166} = 23.8$, $p = 1.15 \times 10^{-6}$) and general proneness ($F_{1,86} = 10.5$, $p = 1.22 \times 10^{-3}$) measures.

These results suggest that while both recent and general hallucination proneness correlate with CH rates, recent hallucination status most strongly relates to frequency of hallucinations and general proneness has a stronger relationship with how the participant learns over time. To further disambiguate these effects, we fit behavioral data to a computational model describing latent states driving task performance.

*High prior precision and decision noise are related to hallucination severity*

We fit the parameters of a three-tiered HGF model using behavior on the CH task (**Fig 4a**). We found that higher CH rates correlated with greater weighting of prior expectations relative to incoming sensory evidence ($\nu$; $t_{86} = 9.26$, $p = 1.47 \times 10^{-15}$, $R = .702$). $\nu$ also negatively correlated with confidence when reporting "no" on no-voice ($F_{1,86} = 8.24$, $p = 5.16 \times 10^{-3}$, $R = -.30$), 25% ($F_{1,86} = 6.75$, $p = 0.01$, $R = -.27$) and 50% trials ($F_{1,85} = 4.55$, $p = 0.036$, $R = -.23$) condition and when saying "yes" on 75% condition ($F_{1,86} = 5.64$, $p = 0.02$, $R = -.25$).

Those in the AH+ group relied more on priors ($H_1 = 4.91$, $p = 0.03$; **Fig 4c**) and reliance on priors increased as severity of recent hallucinations increased ($t_{86} = 2.89$, $p = 4.83 \times 10^{-3}$, $R = .280$; **Fig S2a**). However, we did not find a relationship between prior over-weighting and general hallucination proneness (Low- and High-Propensity: $H_1 = 2.13$, $p = .145$ ; LSHS Total Score: $t_{86} = 0.78$, $p = .439$; **Fig 4f** & **S2c**). We additionally found that those who had higher hallucination proneness had significantly higher decision noise ($\beta^{-1}$; Low- and High-Propensity: $H_1 = 8.61$, $p = 3.34 \times 10^{-3}$; **Fig 4g**) but not seen when comparing groups based on recent hallucinations (AH- and AH+: $H_1 = 2.61$, $p = 8.44 \times 10^{-3}$; **Fig 4d**). However, decision temperature increased with both general (LSHS Total Score: $F_{1,86} = 8.57$, $p = 4.38 \times 10^{-3}$, $R = .28$; **Fig S2d**) and recent hallucination severity scores ($F_{1,86} = 7.27$, $p = 8.44 \times 10^{-3}$, $R = .31$; **Fig S2b**). In general, belief-state trajectories did not differ between groups (**Fig 4b, 4e**); however, we did find a significant effect of recent hallucination group on trial-wise beliefs (**Fig 4b, bottom,** $F_{1,86} = 4.26$, $p = .04$).

Results of metacognitive analyses are available in **Table S9**.

**Discussion**

The primary purpose of this study was to determine if hallucination proneness is related to a tendency to overweight one's perceptual priors as measured by the CH task in the general population. We found increased CH rates in participants with increased hallucination-proneness (as measured by the LSHS) and more severe recent hallucinations (as measured by the AHRS), consistent with past work demonstrating a tendency of individuals with hallucinations to over-rely on priors in clinical and targeted non-clinical samples[5,12,13,44,45] and with continuum models of psychosis[46,47].

Some of these results may provide interesting mechanistic clues about the development of hallucinations. As we and others have argued, a relative over-weighting of priors may be the result of primary or secondary processes, the latter being viewed as a compensatory response to unreliable or noisy incoming sensory evidence[8]. In this general population sample, we again demonstrated a link between higher relative prior weighting ($\nu$) and increased recent hallucination severity scores at the time of assessment. Because this term is defined relative to the weighting afforded to sensory evidence, $\nu$ alone cannot tell us about the absolute weighting of priors or sensory evidence in isolation. However, other results provide some clues as to the status of both priors and sensory precision.

Participants with higher hallucination-proneness have an increased QUEST-derived threshold for initial detection of the most audible stimulus, suggesting lower sensitivity to sensory evidence. Second, those with higher hallucination proneness demonstrate reduced detection rates for more easily audible stimuli. Beyond simple detection, participants with higher hallucination-proneness have reduced confidence in their responses about stimulus detection. These results are consistent with a well-established literature demonstrating sensory deficits in schizophrenia (see [51] for review), and suggest reduced sensory precision–a reduced capacity to effectively use stimuli present in the outside world during perceptual decision-making. This reduced capacity could be due to reduced precision engendered by neurobiological changes such as white matter deterioration, which can be present in people at risk for psychosis[52,53].

Some results also hint at increased absolute prior precision in voice-hearers: high-proneness and AH+ groups begin and end the experiment with higher CH rates than low-proneness and AH- groups (**Fig. 2**), providing evidence for both higher prior precision at the start of the experiment and slower belief updating with new evidence. These differences are also reflected in slower belief updating on HGF analysis (**Fig. 4**), although primarily for AH+/AH- groups. These groups also report proportionally higher rates of detection when stimuli become more difficult to detect, indicating that their perceptual decision-making is more readily driven by expectations than their non-hallucinating counterparts.

It is possible that a general tendency towards over-reporting of detection may be the driver of increased CH rates in the higher-proneness groups. However, this appears unlikely: the higher-proneness group was calculated to have a *higher* threshold using QUEST; over-reporting of detection would have led to a lower calculated threshold, as the adaptive procedure decreases stimulus intensity with each detection reported. It is also possible that this higher threshold estimate could drive a general tendency toward over-reporting detection in the main experiment. However, we do not see a pattern of overall over-endorsement across trial types in the high-proneness group; rather they have *reduced* detection rates on the task with more audible stimuli.

Some of the findings, such as overall reduced confidence, increased detection thresholds, and noisier decision-making in the high-proneness groups, differ from results derived from previously-published, tone-based versions of the CH task[5,13]. Without a direct comparison between the two, it is difficult to say whether these differences in performance are due to the nature of the stimuli being used, but such a difference could point to interesting new directions for research: it may be that different subgroups of psychosis proneness (and psychosis) exist, in which different levels of the sensory processing hierarchy are affected. Alternatively, it may also be possible that different deficits are dominant at different stages of illness, with hallucination-prone groups presumably most closely resembling those in the earliest illness stages. In future planned work, we will compare these two tasks (e.g., linguistic and tone) in the same groups to better assess these differences; we also discuss the importance of longitudinal studies below. Regarding the observation of higher decision noise in hallucination-prone groups, this may be consistent with recent findings of a link between decision noise and diminished cognitive capacity among non-clinical voice-hearers[54].

Interestingly, despite observed lower confidence, participants with higher hallucination-proneness exhibit *increased* overall metacognitive sensitivity and metacognitive efficiency, as measured by meta-$d'$ and the M-ratio of meta-$d'$/$d'$ (see **Table S9** in data supplement). This suggests that while confidence was lower among participants with high hallucination-proneness, they were able to better discriminate between their correct and incorrect answers in terms of their subjective confidence (i.e., participants were more confident in their correct answers and less confident in their incorrect answers). While this finding may seem counterintuitive, when examining the M-ratio at each stimulus threshold, we found that at the 75% condition (i.e., the condition with the most robust sensory evidence available), participants with high hallucination proneness exhibited an M-ratio that exceeded a value of 1, or the "optimal" value of metacognitive efficiency [48], suggesting that individuals with high hallucination-proneness have increased metacognitive awareness (meta-$d'$) as compared to their accuracy on the task ($d'$). Prior work has theorized that individuals who exhibit greater metacognitive awareness compared to task accuracy may be utilizing information other than the available sensory evidence, possibly including heuristics [49], or engaging in post-decisional processes that may affect metacognition [50]. These findings may suggest that rather than an "enhanced" use of sensory information to inform metacognitive awareness during the CH task, utilization of sensory evidence may be altered or supplemented by extraneous information or decisional processes outside of the confines of the task that is not apparent in individuals with low hallucination-proneness.

There are some limitations to the present work. There is a tradeoff between the convenience of a Mechanical-Turk-derived sample and the limitations of accurate self-report of clinical and symptom history [55–57]. This is potentially relevant, as there were a number of participants who had low AHRS and LSHS scores and yet had high CH rates, which were outliers compared to the other participants in the low CH group. It is possible that these participants are simply prone to high CH rates via a different mechanism than hallucination-proneness, that they did not respond to the test accurately in a manner our quality control does not measure, or that they actually were in the high-proneness groups but did not provide accurate self-report data. In addition, this experiment did not record any neural correlates of the behavior observed, limiting our ability to replicate hypothesized links to neurobiology. Future work could be undertaken to correlate these behavioral results with neural measures. For example, proposed correlates of reduced sensory precision could be related to measures of white and gray matter integrity as well as functional connectivity using structural and functional imaging.

Taken together, our results demonstrate that hallucination-proneness and hallucination severity are related to susceptibility to conditioned hallucinations and attendant prior hyper-precision in the general population. Because of the sensitivity of the CH task in this population, similar tools may be developed to screen for or investigate factors driving subclinical psychosis-like phenomena in broad, diverse samples. We also demonstrate for the first time that hallucination-prone individuals may exhibit sensory disturbances that are potentially causally related to their tendency to over-trust their priors. If these results are confirmed and extended using novel paradigms, they may help inform causal models of the development of psychotic symptoms.

**Table 1. Demographic and Clinical Information by Group**

|  | Group by AHRS Score | | Group by LSHS Score | |
|---|---|---|---|---|
|  | AH- | AH+ | Low Propensity | High Propensity |
| n (%) | 45 (51.1) | 43 (48.9) | 73 (83.0) | 15 (17.0) |
| Age mean (sd) | **42.0 (10.8)** | **36.1 (10.1)** | **40.9 (10.9)** | **30.5 (4.3)** |
| Sex n (%M) | 29 (64.4) | 33 (76.7) | 48 (65.7) | 14 (93.3) |
| Race n (%) |  |  |  |  |
| American Indian/Alaska Native | 0 (0.0) | 3 (6.9) | 2 (2.7) | 1 (6.7) |
| Asian | 5 (11.1) | 5 (11.6) | 8 (11.0) | 2 (13.3) |
| Black or African American | 1 (2.2) | 4 (9.3) | 4 (5.5) | 1 (6.7) |
| White | 36 (80.0) | 31 (72.1) | 57 (78.1) | 10 (66.7) |
| More than one race | 3 (6.7) | 0 (0.0) | 2 (2.7) | 1 (6.7) |
| Mental Illness n (%) | 4 (8.9) | 1 (2.3) | 5 (6.8) | 0 (0.0) |
| Medication Current n (%) | 1 (2.2) | 2 (4.7) | 2 (2.7) | 1 (6.6) |
| LSHS mean (sd) | **2.2 (3.7)** | **12.3 (10.8)** | **3.3 (4.0)** | **25.8 (3.9)** |
| AHRS mean (sd) | **0.0 (0.0)** | **9.8 (7.5)** | 2.5 (4.2) | 16.0 (8.3) |
| ACE mean (sd) | 1.2 (1.9) | 0.93 (1.3) | 1.2 (1.7) | 0.7 (1.2) |
| THQ mean (sd) | **2.2 (2.1)** | **1.1 (1.6)** | **1.9 (2.0)** | **0.20 (0.77)** |
| PHQ-9 mean (sd) | **2.4 (3.2)** | **6.4 (6.6)** | **3.0 (4.3)** | **10.8 (6.2)** |
| GAD-7 mean (sd) | **1.6 (2.5)** | **4.0 (5.0)** | **2.0 (3.5)** | **6.3 (4.9)** |
| PDI Total mean (sd) | 1.6 (2.4) | 1.5 (2.0) | 1.4 (2.0) | 2.3 (3.3) |

**Figures & Figure Legends**

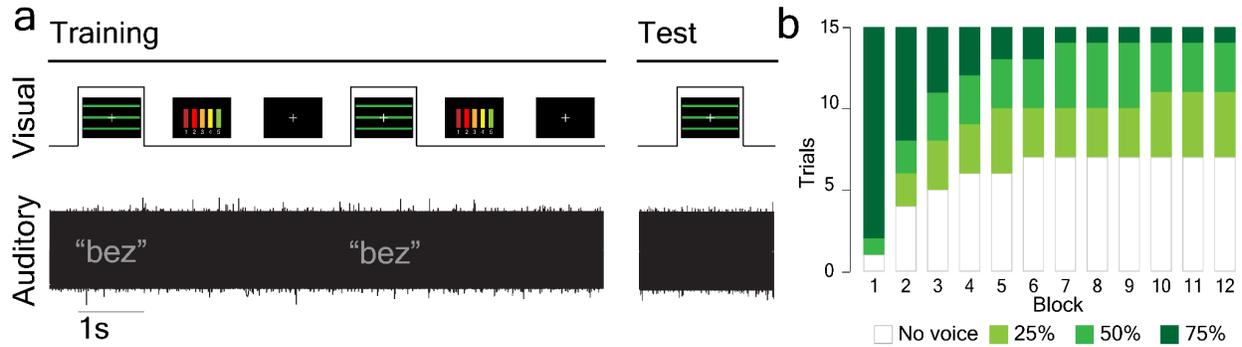

**Figure 1. Conditioned Hallucination (CH) Task Structure. a.** One of three visual patterns and voice pairs were presented simultaneously. White noise played throughout the task. After stimuli were presented, participants indicated by button-press whether they heard the voice or not and then rated confidence in their decision. **b.** Using the QUEST thresholding procedure, we estimated 75-, 50-, and 25-percent detection thresholds for each participant. Presentation of stimuli at each intensity level was systematically varied over the course of 12 blocks of 15 trials each, with 75-percent trials becoming more infrequent and quieter and absent-tone trials becoming more frequent.

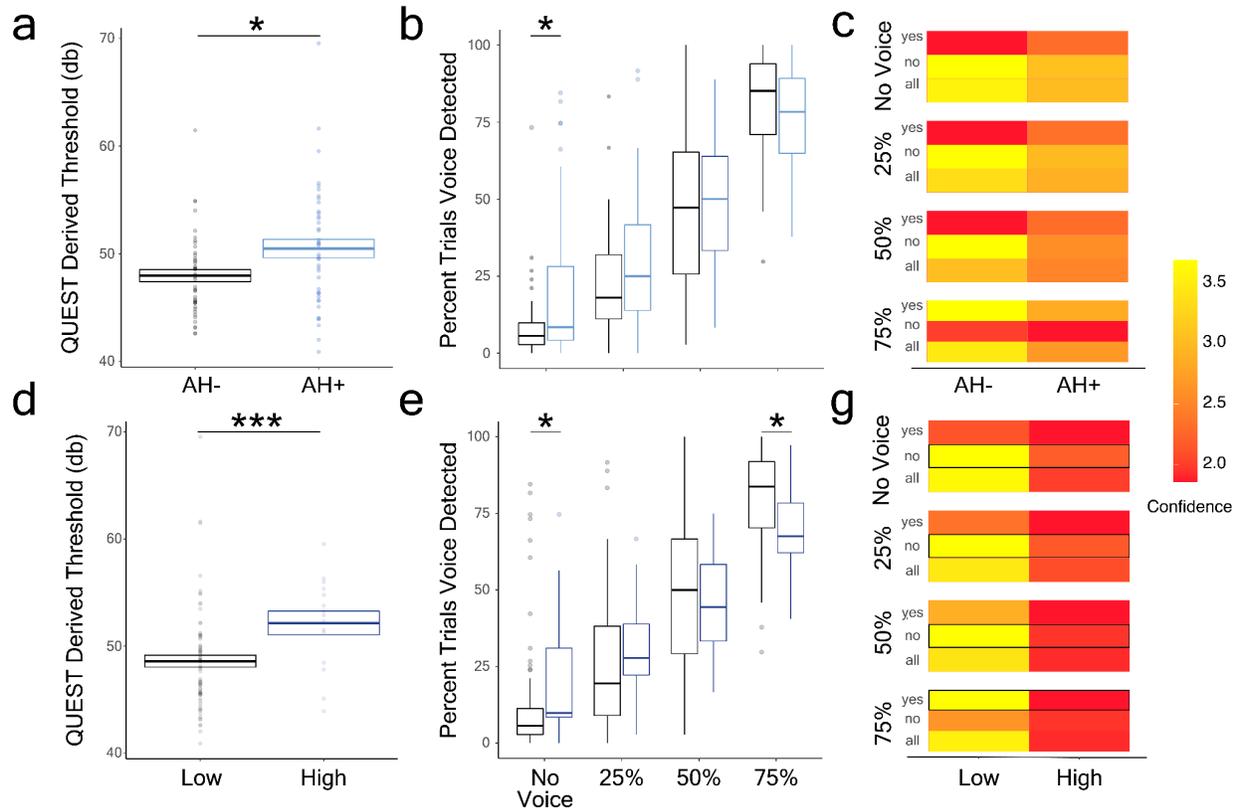

**Figure 2. Hallucination proneness is associated with poorer sensory performance and greater CH rates. a., d.** Estimated 75% detection thresholds. Estimated thresholds were higher for those who reported any hallucinations in the last 24 hours (AH+, **a**) and those with high general propensity for hallucinations (High-Propensity, **d**). **b., e.** Percent of reported voice-detected trials for No-Voice and at each estimated threshold. **c., f.** Mean confidence for each response by target volume and response. Black boxes indicate significant differences. In general, hallucination-prone groups were less confident in their decisions; however, confidence was not significantly different for any given condition between AH- and AH+ groups. High-proneness groups were significantly less confident when reporting no on No-Voice, 25, and 50% and when reporting yes on 75% conditions. *, $p < 0.05$; ***, $p < 0.001$

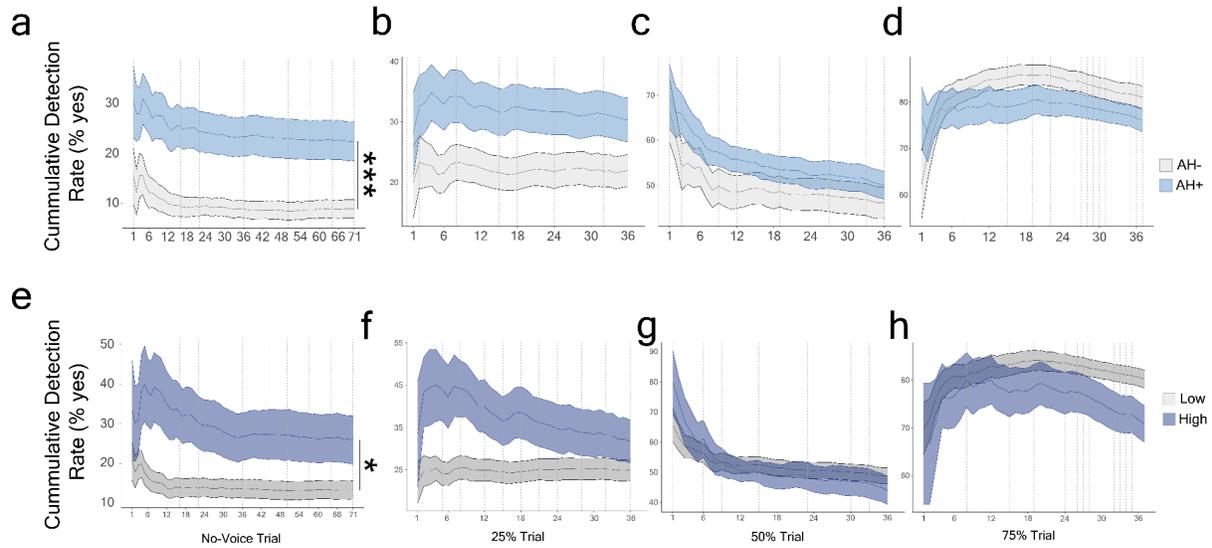

**Figure 3. High-proneness and high-severity groups maintain high CH rates over time. a., e.** Cumulative CH rates. Both AH+ and High-Propensity groups were more likely to report hearing a voice in No-Voice trials. There was also a significant interaction with block, indicating that groups differed in learning rates. This trend was observed in the 25-percent conditions (**b, f**), although it did not reach statistical significance. Behavior very similar comparable between groups for 50-percent trials (**c, g**). On 75-percent trials **(d, h)**, High Proneness participants were significantly less likely to report hearing the voice, but this difference was not evident between AH- and AH+ groups. There was a significant block-by-group interaction for both groupings. *, $p < 0.05$; ** $p < 0.01$

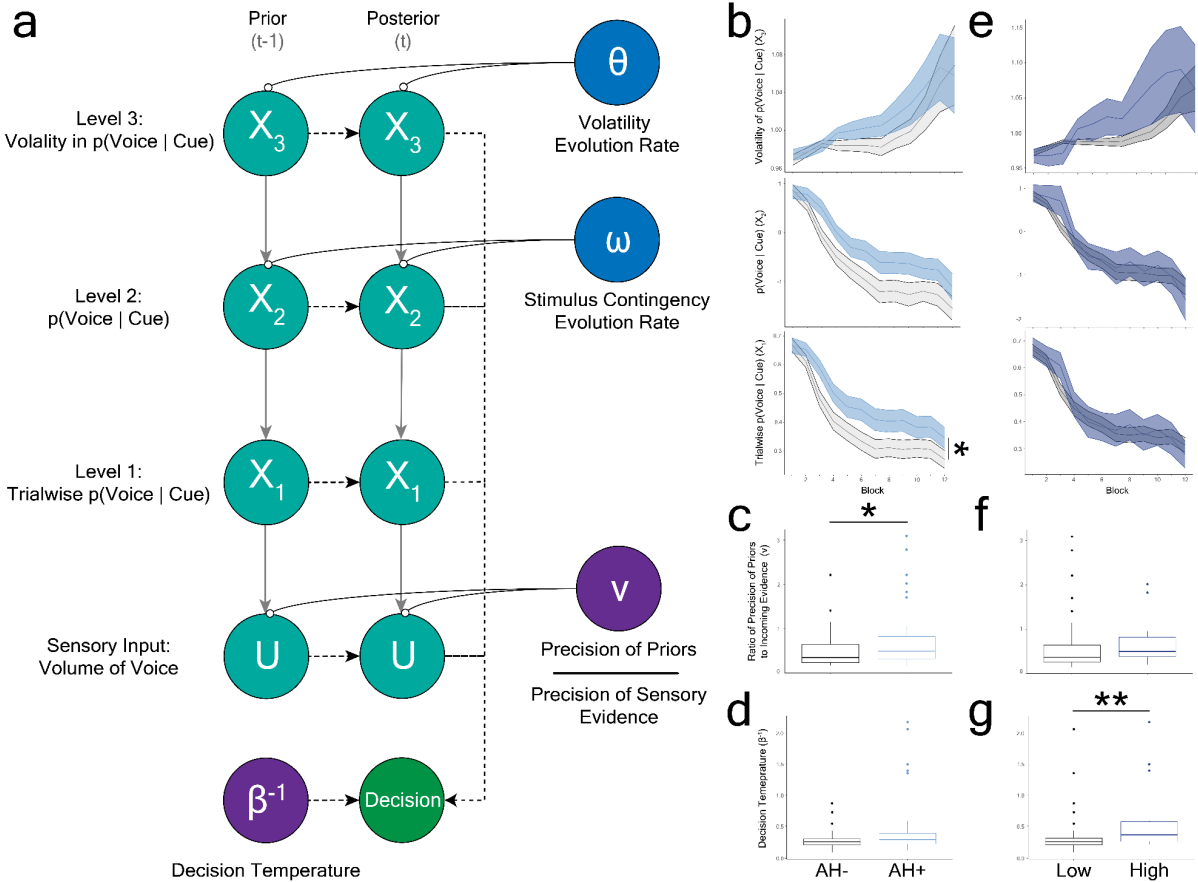

**Figure 4. High prior precision and decision noise are related to hallucination severity on Hierarchical Gaussian Filter (HGF) analysis. a.** The HGF model uses stimulus strength (U) and responses to estimate latent states driving behavior on the linguistic conditioned hallucination task. These states include: belief trajectories at three levels ($X_1$, $X_2$, $X_3$), the evolution rates of these beliefs ($\omega$ and $\theta$), relative precision of prior ($v$), and decision noise ($\beta^{-1}$). $X_1$ corresponds with trial-wise probability that a voice was present. $X_2$ reflects the global probability that there is a voice when the pattern is present. $X_3$ is belief in the volatility in the association between the voice and pattern. For those who had experienced hallucinations within the last 24 hours (**b-d**), we did not find significant differences in belief-states at $X_3$ (**b, top**), or $X_2$ (**b, middle**). AH+ had stronger trial-wise beliefs that a voice was present (**b, bottom**) and weighted priors more heavily (**c**). However AH- and AH+ groups did not differ in decision noise ($\beta^{-1}$; **d**). For groups reflecting general hallucination severity (**e-g**), They did not significantly differ in belief states (**e**) nor how heavily they weighted their priors (**f**). However, they did have higher decision noise (**g**).

**References**

1.  Benrimoh D, Parr T, Vincent P, Adams RA, Friston K. Active Inference and Auditory Hallucinations. *Comput Psychiatr*. 2018;2:183-204.

2.  Benrimoh D, Parr T, Adams RA, Friston K. Hallucinations both in and out of context: An active inference account. *PLoS One*. 2019;14(8):e0212379.

3.  Benrimoh D, Tanguay-Sela M, Perlman K, et al. Using a simulation centre to evaluate preliminary acceptability and impact of an artificial intelligence-powered clinical decision support system for depression treatment on the physician–patient interaction. *BJPsych Open*. 2021;7(1). doi:10.1192/bjo.2020.127

4.  Chen AG, Benrimoh D, Parr T, Friston KJ. A Bayesian Account of Generalist and Specialist Formation Under the Active Inference Framework. *Frontiers Artificial Intelligence Appl*. 2020;3. doi:10.3389/frai.2020.00069

5.  Powers AR, Mathys C, Corlett PR. Pavlovian conditioning-induced hallucinations result from overweighting of perceptual priors. *Science*. 2017;357(6351):596-600.

6.  Powers AR. Mathematical nosology: Computational approaches to understanding psychosis. *Schizophr Res*. 2022;245:1-4.

7.  Fisher VL, Ortiz LS, Powers AR 3rd. A computational lens on menopause-associated psychosis. *Front Psychiatry*. 2022;13:906796.

8.  Sheldon AD, Kafadar E, Fisher V, et al. Perceptual pathways to hallucinogenesis. *Schizophr Res*. Published online February 22, 2022. doi:10.1016/j.schres.2022.02.002

9.  Adams RA, Vincent P, Benrimoh D, Friston KJ, Parr T. Everything is connected: Inference and attractors in delusions. *Schizophr Res*. Published online August 9, 2021. doi:10.1016/j.schres.2021.07.032

10. Corlett PR, Horga G, Fletcher PC, Alderson-Day B, Schmack K, Powers AR 3rd. Hallucinations and Strong Priors. *Trends Cogn Sci*. 2019;23(2):114-127.

11. Friston K. A theory of cortical responses. *Philos Trans R Soc Lond B Biol Sci*. 2005;360(1456):815-836.

12. Kafadar E, Mittal VA, Strauss GP, et al. Modeling perception and behavior in individuals at clinical high risk for psychosis: Support for the predictive processing framework. *Schizophr Res*. 2020;226:167-175.

13. Kafadar E, Fisher V, Quagan B, et al. Conditioned hallucinations and prior over-weighting are state-sensitive markers of hallucination susceptibility. *Biol Psychiatry*. Published online May 13, 2022. doi:10.1016/j.biopsych.2022.05.007

14. Teufel C, Subramaniam N, Dobler V, et al. Shift toward prior knowledge confers a perceptual advantage in early psychosis and psychosis-prone healthy individuals. *Proc Natl Acad Sci U S A*. 2015;112(43):13401-13406.

15. Vercammen A, Aleman A. Semantic expectations can induce false perceptions in hallucination-prone individuals. *Schizophr Bull*. 2010;36(1):151-156.

16. Friston K, Kilner J, Harrison L. A free energy principle for the brain. *J Physiol Paris*.

2006;100(1-3):70-87.

17. Friston K, Kiebel S. Predictive coding under the free-energy principle. *Philos Trans R Soc Lond B Biol Sci*. 2009;364(1521):1211-1221.

18. Noppeney U, Josephs O, Hocking J, Price CJ, Friston KJ. The effect of prior visual information on recognition of speech and sounds. *Cereb Cortex*. 2008;18(3):598-609.

19. Lau EF, Phillips C, Poeppel D. A cortical network for semantics: (de)constructing the N400. *Nat Rev Neurosci*. 2008;9(12):920-933.

20. Lee TS, Mumford D. Hierarchical Bayesian inference in the visual cortex. *J Opt Soc Am A Opt Image Sci Vis*. 2003;20(7):1434-1448.

21. Petzschner FH, Glasauer S, Stephan KE. A Bayesian perspective on magnitude estimation. *Trends Cogn Sci*. Published online 2015. doi:10.1016/j.tics.2015.03.002

22. Dayan P. A hierarchical model of binocular rivalry. *Neural Comput*. 1998;10(5):1119-1135.

23. Schellekens W, van Wezel RJ, Petridou N, Ramsey NF, Raemaekers M. Predictive coding for motion stimuli in human early visual cortex. *Brain Struct Funct*. Published online 2014. doi:10.1007/s00429-014-0942-2

24. Baldeweg T. Repetition effects to sounds: evidence for predictive coding in the auditory system. *Trends Cogn Sci*. 2006;10(3):93-94.

25. Benrimoh D, Sheldon A, Sibarium E, Powers AR. Computational Mechanism for the Effect of Psychosis Community Treatment: A Conceptual Review From Neurobiology to Social Interaction. *Front Psychiatry*. 2021;12:685390.

26. Abdul-Rahman MF, Qiu A, Woon PS, Kuswanto C, Collinson SL, Sim K. Arcuate fasciculus abnormalities and their relationship with psychotic symptoms in schizophrenia. *PLoS One*. 2012;7(1):e29315.

27. Higley MJ, Picciotto MR. Neuromodulation by acetylcholine: examples from schizophrenia and depression. *Curr Opin Neurobiol*. 2014;29:88-95.

28. Cella M, Cooper A, Dymond SO, Reed P. The relationship between dysphoria and proneness to hallucination and delusions among young adults. *Compr Psychiatry*. 2008;49(6):544-550.

29. Verdoux H, van Os J, Maurice-Tison S, Gay B, Salamon R. Increased occurrence of depression in psychosis-prone subjects. A follow-up study in primary care settings. *Schizophrenia Research*. 2000;41(1):76. doi:10.1016/s0920-9964(00)90478-x

30. Stainsby LM, Lovell GP. Proneness to hallucinations and delusions in a non‑clinical sample: Exploring associations with metacognition and negative affect. *Aust J Psychol*. 2014;66(1):1-7.

31. Kroenke Kurt, Spitzer Robert L. The PHQ-9: A New Depression Diagnostic and Severity Measure. *Psychiatr Ann*. 2002;32(9):509-515.

32. Spitzer RL, Kroenke K, Williams JBW, Löwe B. A brief measure for assessing generalized

anxiety disorder: the GAD-7. *Arch Intern Med*. 2006;166(10):1092-1097.

33. Siddi S, Ochoa S, Laroi F, et al. A Cross-National Investigation of Hallucination-Like Experiences in 10 Countries: The E-CLECTIC Study. *Schizophr Bull*. 2019;45(45 Suppl 1):S43-S55.

34. Hoffman RE, Hawkins KA, Gueorguieva R, et al. Transcranial magnetic stimulation of left temporoparietal cortex and medication-resistant auditory hallucinations. *Arch Gen Psychiatry*. 2003;60(1):49-56.

35. Hooper LM, Stockton P, Krupnick JL, Green BL. Development, Use, and Psychometric Properties of the Trauma History Questionnaire. *J Loss Trauma*. 2011;16(3):258-283.

36. Felitti VJ, Anda RF, Nordenberg D, et al. Relationship of childhood abuse and household dysfunction to many of the leading causes of death in adults: The Adverse Childhood Experiences (ACE) study. *Am J Prev Med*. 56(6):774-786.

37. Watson AB, Pelli DG. QUEST: a Bayesian adaptive psychometric method. *Percept Psychophys*. 1983;33(2):113-120.

38. Mathys CD, Lomakina EI, Daunizeau J, et al. Uncertainty in perception and the Hierarchical Gaussian Filter. *Front Hum Neurosci*. 2014;8:825.

39. Mathys C, Daunizeau J, Friston KJ, Stephan KE. A bayesian foundation for individual learning under uncertainty. *Front Hum Neurosci*. 2011;5:39.

40. Stephan KE, Penny WD, Daunizeau J, Moran RJ, Friston KJ. Bayesian model selection for group studies. *Neuroimage*. 2009;46(4):1004-1017.

41. NparLD: Nonparametric tests for repeated measures data in factorial... In nparLD: Nonparametric analysis of longitudinal data in factorial experiments. Published August 7, 2022. Accessed June 14, 2023. https://rdrr.io/cran/nparLD/man/nparLD.html

42. The R Project for Statistical Computing. Accessed January 9, 2023. https://www.R-project.org/

43. Gizdic A, Baxter T, Barrantes-Vidal N, Park S. Loneliness and psychosocial predictors of psychosis-proneness during COVID-19: Preliminary findings from Croatia. *Psychiatry Res*. 2022;317:114900.

44. Alderson-Day B, Lima CF, Evans S, et al. Distinct processing of ambiguous speech in people with non-clinical auditory verbal hallucinations. *Brain*. 2017;140:2475-2489.

45. Cassidy CM, Balsam PD, Weinstein JJ, et al. A Perceptual Inference Mechanism for Hallucinations Linked to Striatal Dopamine. *Curr Biol*. 2018;28(4):503-514 e4.

46. van Os J, Linscott RJ, Myin-Germeys I, Delespaul P, Krabbendam L. A systematic review and meta-analysis of the psychosis continuum: evidence for a psychosis proneness-persistence-impairment model of psychotic disorder. *Psychol Med*. 2009;39(2):179-195.

47. Linscott RJ, van Os J. Systematic reviews of categorical versus continuum models in psychosis: evidence for discontinuous subpopulations underlying a psychometric continuum. Implications for DSM-V, DSM-VI, and DSM-VII. *Annu Rev Clin Psychol*. 2010;6:391-419.

48. Fleming SM, Lau HC. How to measure metacognition. *Front Hum Neurosci*. 2014;8:443.

49. Maniscalco B, Peters MAK, Lau H. Heuristic use of perceptual evidence leads to dissociation between performance and metacognitive sensitivity. *Atten Percept Psychophys*. 2016;78(3):923-937.

50. Moreira CM, Rollwage M, Kaduk K, Wilke M, Kagan I. Post-decision wagering after perceptual judgments reveals bi-directional certainty readouts. *Cognition*. 2018;176:40-52.

51. Dondé C, Avissar M, Weber MM, Javitt DC. A century of sensory processing dysfunction in schizophrenia. *Eur Psychiatry*. 2019;59:77-79.

52. Straub KT, Hua JPY, Karcher NR, Kerns JG. Psychosis risk is associated with decreased white matter integrity in limbic network corticostriatal tracts. *Psychiatry Res Neuroimaging*. 2020;301:111089.

53. de Weijer AD, Neggers SFW, Diederen KMS, et al. Aberrations in the arcuate fasciculus are associated with auditory verbal hallucinations in psychotic and in non-psychotic individuals. *Hum Brain Mapp*. 2013;34(3):626-634.

54. Baumeister D, Sedgwick O, Howes O, Peters E. Auditory verbal hallucinations and continuum models of psychosis: A systematic review of the healthy voice-hearer literature. *Clin Psychol Rev*. 2017;51:125-141.

55. Burnette CB, Luzier JL, Bennett BL, et al. Concerns and recommendations for using Amazon MTurk for eating disorder research. *Int J Eat Disord*. 2022;55(2):263-272.

56. Kan IP, Drummey AB. Do imposters threaten data quality? An examination of worker misrepresentation and downstream consequences in Amazon's Mechanical Turk workforce. *Comput Human Behav*. 2018;83:243-253.

57. Hauser D, Paolacci G, Chandler JJ. Common Concerns with MTurk as a Participant Pool: Evidence and Solutions. doi:10.31234/osf.io/uq45c

58. Rounis E, Maniscalco B, Rothwell JC, Passingham RE, Lau H. Theta-burst transcranial magnetic stimulation to the prefrontal cortex impairs metacognitive visual awareness. *Cogn Neurosci*. 2010;1(3):165-175.

59. Maniscalco B, Lau H. A signal detection theoretic approach for estimating metacognitive sensitivity from confidence ratings. *Conscious Cogn*. 2012;21(1):422-430.

60. Palmer EC, David AS, Fleming SM. Effects of age on metacognitive efficiency. *Conscious Cogn*. 2014;28:151-160.

61. Charles L, Van Opstal F, Marti S, Dehaene S. Distinct brain mechanisms for conscious versus subliminal error detection. *Neuroimage*. 2013;73:80-94.

**Supplemental Materials**

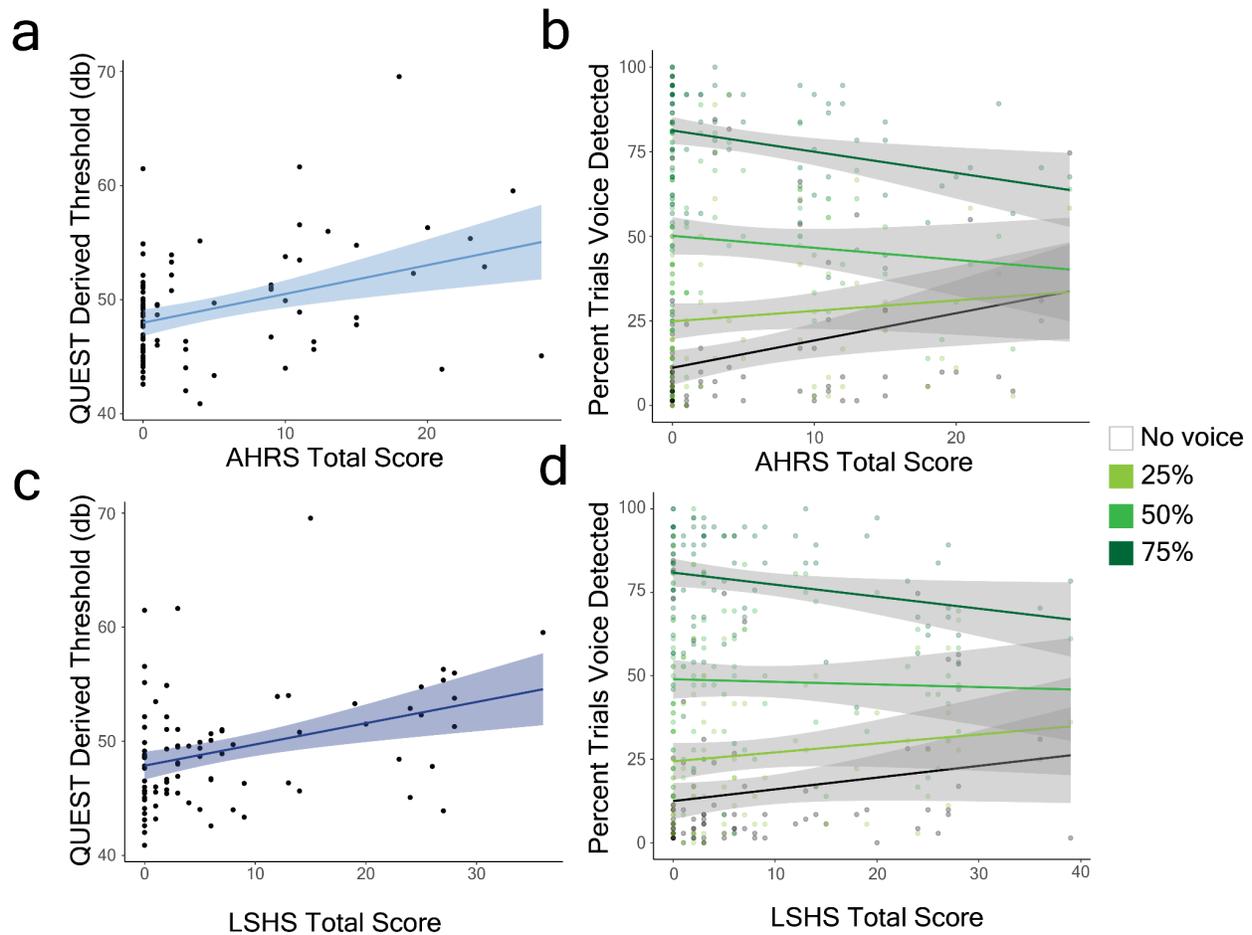

**Figure S1. Regression Analysis of Behavioral Data and Hallucination-Severity Scores.**
QUEST-estimated 75% threshold compared to AHRS (**a**) and LSHS (**c**) total scores. Threshold volume significantly increased as both recent and general severity of hallucinations increased. **b., d.** Detection rates for each presented condition. For 75% and 50% trials, those with higher hallucination severity were less likely to report detecting the tone. For No-voice and 25% trials, hallucination-prone individuals were more likely to report hearing the tone. Of these relationships, we found significant effects for 75% and No-voice with both AHRS and LSHS total score.

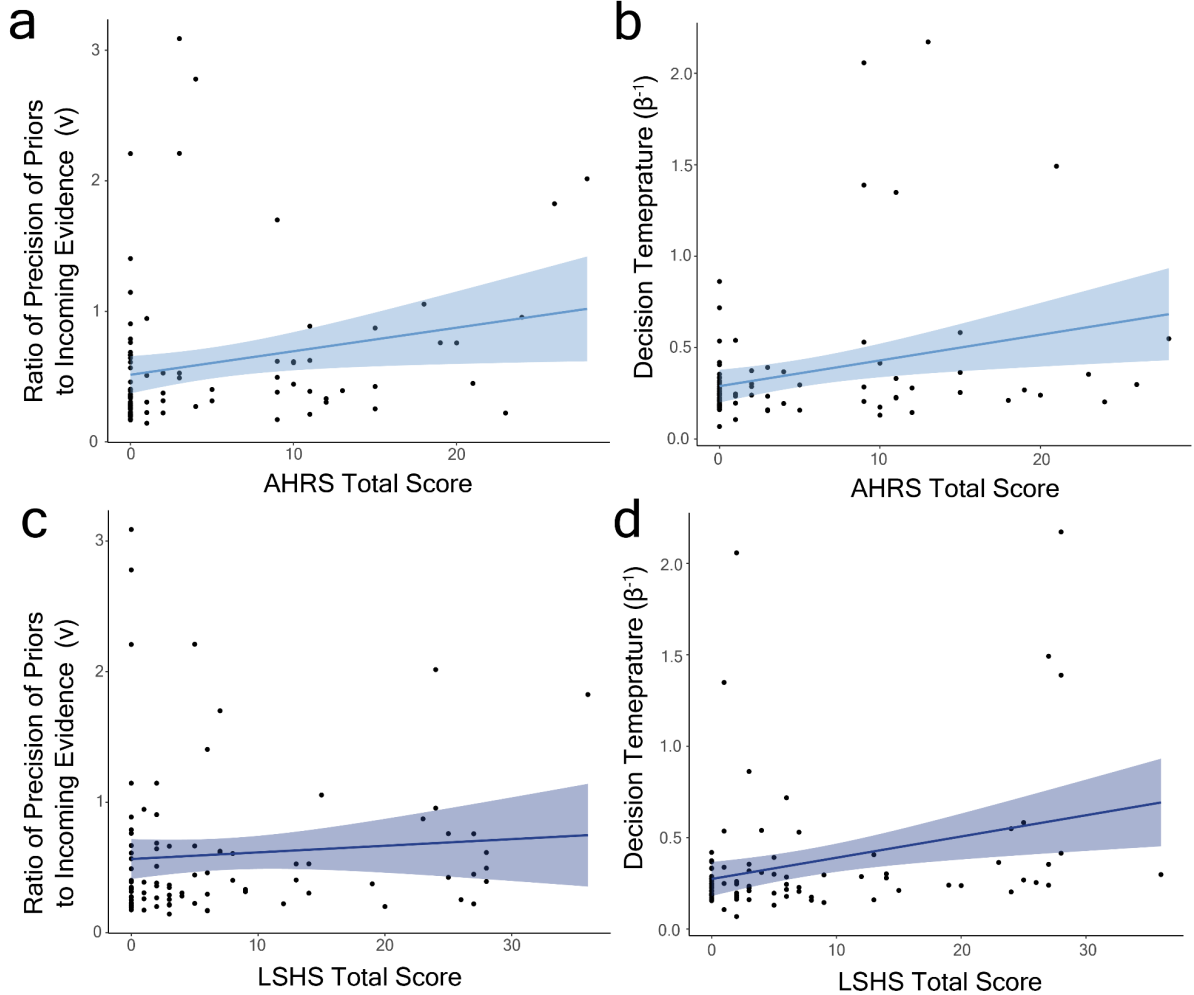

**Figure S2. Regression Analysis of HGF Model Estimates of Latent States and Hallucination-Severity Scores.** Consistent with group differences (**Fig 5**), prior-precision increased with severity of recent hallucinations (**a**) but not general (**c**). For both severity measures decision temperature / noise increased with severity score (**b,d**).

**Table S1. Endorsed Psychiatric Illnesses in Sample**

|  | AH- | AH+ | Low-Propensity | High-Propensity |
|---|---|---|---|---|
| Depression n(%) | 3 (6.7) | 1 (2.3) | 4 (5.5) | 0 (0.0) |
| Anxiety* n(%) | 2 (4.4) | 0 (0.0) | 2 (2.7) | 0 (0.0) |
| Obsessive Compulsive Disorder n(%) | 1 (2.2) | 0 (0.0) | 1 (1.4) | 0 (0.0) |

* Includes social and general anxiety

**Table S2. Results from Repeated Measures ANOVA of Confidence Across Trial and Response Types between AH- and AH+ Groups**

|  | SS | MS | F values | p-value |
|---|---|---|---|---|
| Between |  |  |  |  |
| Response | 2.80 | 1.39 | 0.09 | .92 |
| Condition | 2.60 | 1.32 | 0.08 | .92 |
| AH- and AH+ Group | 4.60 | 4.59 | 0.29 | .59 |
| Resp * Group | 6.50 | 3.27 | 0.21 | .82 |
| Condition * Group | 4.80 | 4.84 | 0.30 | .58 |
| Within |  |  |  |  |
| Response | 65.4 | 32.7 | 132.2 | $3.18 \times 10^{-51}$ |
| Condition | 2.41 | 0.8 | 3.25 | .021 |
| Resp * Condition | 96.51 | 16.08 | 65.07 | $1.86 \times 10^{-67}$ |
| Resp * Group | 15.80 | 7.90 | 31.95 | $3.82 \times 10^{-14}$ |
| Condition * Group | 0.05 | 0.18 | 0.73 | .54 |
| Resp * Cond * Group | 11.81 | 1.97 | 7.96 | $2.18 \times 10^{-8}$ |

**Table S3. Results from Repeated Measures ANOVA of Confidence Across Trial and Response Types between Low- and High-Propensity Groups**

|  | SS | MS | F value | p-value |
|---|---|---|---|---|
| Between |  |  |  |  |
| Response | 2.80 | 1.39 | 0.10 | .90 |
| Condition | 2.60 | 1.32 | 0.10 | .91 |
| Low- and High-Propensity Group | 105.7 | 104.7 | 7.81 | $6.47 \times 10^{-3}$ |
| Resp x Group | 75.90 | 75.87 | 5.61 | 0.02 |
| Within |  |  |  |  |
| Response | 65.4 | 32.7 | 130.2 | $1.55 \times 10^{-50}$ |
| Condition | 2.41 | 0.80 | 3.20 | .02 |
| Resp x Condition | 96.5 | 16.1 | 65.1 | $1.51 \times 10^{-66}$ |
| Resp x Group | 7.48 | 3.74 | 14.9 | $4.27 \times 10^{-7}$ |
| Condition x Group | 0.22 | 0.07 | 0.30 | .83 |
| Resp x Cond x Group | 16.9 | 2.81 | 11.2 | $4.35 \times 10^{-12}$ |

**Table S4. Results from Repeated Measures ANOVA of Cumulative CH Rates Across Trials between AH- and AH+ Groups**

|  | SS | MS | F values | p-value |
|---|---|---|---|---|
| Between |  |  |  |  |
| AH- and AH+ Group | 33.29 | 33.29 | 11.68 | $9.67 \times 10^{-3}$ |
| Within |  |  |  |  |
| Trial | 1.19 | 1.90 | 162.8 | $4.89 \times 10^{-37}$ |
| Group * Trial | 0.06 | 0.056 | 8.20 | $4.20 \times 10^{-3}$ |

**Table S5. Results from Repeated Measures ANOVA of Cumulative CH Rates Across Trials between Low- and High-Propensity Groups**

|  | SS | MS | F values | p-value |
|---|---|---|---|---|
| Between |  |  |  |  |
| Low- and High-Propensity Group | 20.74 | 20.74 | 6.924 | 0.010 |
| Within |  |  |  |  |
| Trial | 1.19 | 1.19 | 165.2 | $2.45 \times 10^{-37}$ |
| Group * Trial | 0.44 | 0.44 | 61.3 | $5.74 \times 10^{-15}$ |

**Table S6. Results from Repeated Measures ANOVA of Cumulative CH Rates Across Trials including both AH- and AH+ and Low- and High-Propensity groups**

|  | SS | MS | F values | p-value |
|---|---|---|---|---|
| Between |  |  |  |  |
| AH- and AH+ Group | 33.29 | 33.29 | 11.8 | $9.12 \times 10^{-4}$ |
| Low- and High-Propensity Group | 5.92 | 5.92 | 2.11 | 0.15 |
| AH- and AH+ * Low- and High-Propensity Groups | 2.69 | 2.69 | 0.96 | 0.33 |
| Within |  |  |  |  |
| Trial | 1.19 | 1.19 | 165.2 | $2.14 \times 10^{-37}$ |
| AH- and AH+ * Trial | 0.06 | 0.06 | 8.29 | $4.01 \times 10^{-3}$ |
| Low- and High-Propensity * Trial | 0.38 | 0.38 | 53.2 | $3.38 \times 10^{-13}$ |
| AH * Low- and High-Propensity Groups * Trial | 0.09 | 0.09 | 12.1 | $4.94 \times 10^{-4}$ |

**Table S7. Results from Repeated Measures ANOVA of Cumulative 75% Detection Rates Across Trials between AH- and AH+ Groups**

|  | SS | MS | F values | p-value |
|---|---|---|---|---|
| Between |  |  |  |  |

| | SS | MS | F values | p-value |
|---|---|---|---|---|
| AH- and AH+ Group | 1.19 | 1.19 | 1.03 | .31 |
| Within | | | | |
| Trial | 0.34 | 0.34 | 27.09 | $2.06 \times 10^{-7}$ |
| Group * Trial | 0.30 | 0.30 | 23.75 | $1.15 \times 10^{-6}$ |

**Table S8. Results from Repeated Measures ANOVA of Cumulative 75% Detection Rates Across Trials between Low- and High-Propensity Groups**

| | SS | MS | F values | p-value |
|---|---|---|---|---|
| Between | | | | |
| LSHS Group | 0.13 | 0.13 | 1.16 | .286 |
| Within | | | | |
| Trial | 0.34 | 0.34 | 26.98 | $2.18 \times 10^{-7}$ |
| Group * Trial | 0.13 | 0.13 | 10.48 | $1.22 \times 10^{3}$ |

**Table S9. Group comparisons of metacognitive sensitivity (meta-*d'*, M-Ratio, and M-Difference) across trial conditions**

| Measure | Condition | Median | | Mann-Whitney | | Median | | Mann-Whitney | |
|---|---|---|---|---|---|---|---|---|---|
| | | AH+ | AH- | U | p-value | High Propensity | Low Propensity | U | p-value |
| Meta-*d'* | Overall | .62 | .62 | 1022.0 | .65 | **.93** | **.48** | **747.0** | **.03** |
| | 75% | 1.48 | 1.85 | 807.0 | .31 | 2.77 | 1.37 | 694.0 | .105 |
| | 50% | 1.43 | 1.66 | 942.0 | .46 | .96 | .50 | 697.0 | .098 |
| | 25% | .58 | .57 | 788.0 | .64 | .41 | .21 | 722.0 | .054 |
| M-Ratio | Overall | .37 | .42 | 922.0 | .71 | **.56** | **.35** | **741.0** | **.03** |
| | 75% | .69 | .84 | 961.0 | .96 | **1.20** | **.69** | **790.0** | **.007** |
| | 50% | .47 | .51 | 1076.0 | .37 | **.89** | **.42** | **8270** | **.002** |
| | 25% | .40 | .43 | 891.0 | .53 | .77 | .40 | 693.0 | .11 |
| M-Diff | Overall | -0.75 | -.78 | 982.0 | 0.91 | -.48 | -.80 | 695.0 | .10 |
| | 75% | -0.84 | -.34 | 777.0 | .11 | .43 | -.47 | 710.0 | .07 |
| | 50% | -.61 | -.44 | 913.0 | .65 | -.04 | -.56 | 652.0 | .25 |
| | 25% | -.15 | -.22 | 1091.0 | .30 | -.09 | -.23 | 596.0 | .59 |

*Exploratory Analyses of Metacognitive Sensitivity and Efficiency*

Given the general reduction of confidence observed among the hallucination-prone group, we performed exploratory analyses of metacognitive sensitivity as ascertained by meta-$d'$ [58,59]. In brief, meta-$d'$ is a bias-free measure of metacognition as derived from a Signal Detection Theory (SDT) framework (i.e., the amount of signal available for metacognitive processing secondary to the first-order task of stimulus classification). Meta-$d'$ is an analogue to the classical sensitivity index known as $d'$ ( Meta-$d'$ reflects a participant's ability to discriminate between correct and incorrect responses in terms of subjective confidence while also controlling for overall task performance as determined by $d'$. In addition to meta-$d'$, because both $d'$ and meta-$d'$ are calculated in the same signal-to-noise ratio units of measurement, the two measures can be directly compared. We computed meta-$d'$, an M-ratio (meta-$d'$/$d'$), a measure of metacognitive efficiency, and an M-difference score (meta-$d'$ – $d'$) for each participant at each condition level (overall, 75%, 50%, and 25%) to compare metacognitive efficiency between groups. We did not calculate these metrics at the 0% condition, as mathematically, this is the negation of overall meta-$d'$ (i.e., meta-$d'$ of all no-signal trials as compared to all signal trials). The M-ratio can be conceptualized as the amount of sensory information (as determined by $d'$) that is utilized when making metacognitive judgements, with 'optimal' metacognitive sensitivity being an M-ratio of 1 [48]. An M-ratio greater than 1 suggests that awareness is more precise than accuracy, and this may be reflective of subjects' engagement of post-decisional processes, use of information outside of the task paradigm (e.g., heuristics), or when decisions are made under time pressure [60,61]. Relatedly, the M-difference score is distance in terms of maximum utility of sensory information of meta-$d'$ and $d'$, with 'optimal' M-difference being 0 [48].

In our modeling of meta-$d'$, we found no significant group differences in metacognitive sensitivity or efficiency at any condition between the AH+ and AH- groups (**Table S9).** However, when comparing the Low and High hallucination-prone groups, we found significant group differences in overall meta-$d'$ between the Low- and High-Propensity groups with the High group showing increased metacognitive sensitivity as compared to the Low-Propensity group. Moreover, we found that compared to individuals in the Low-Propensity group, individuals with high hallucination proneness exhibited increased metacognitive efficiency as reflected by M-ratios, both overall and at the 75% and 50% conditions. We did not observe any group differences in M-difference scores at any condition between the Low- and High-Propensity groups.